\documentclass[manuscript]{aastex}
\usepackage{amsmath, amsthm, amssymb}
\usepackage[T1]{fontenc}
\usepackage{float}
\newcommand{\fom}{f_\Omega}
\newcommand{\edot}{\dot{E}}

\shorttitle{Experimental Constraints on $\gamma$-ray Pulsar Gap Models and the Pulsar GeV to Pulsar Wind Nebula TeV Connection.}
\shortauthors{Abeysekara et al.}

\begin{document}

\title{Experimental Constraints on $\gamma$-ray Pulsar Gap Models and the Pulsar GeV to Pulsar Wind Nebula TeV Connection}

\author{A. U. Abeysekara}
\email{udaraabeysekara@yahoo.com}

\and

\author{J. T. Linnemann}
\affil{Department of Physics and Astronomy, Michigan State University, 567 Wilson Road, East Lansing, MI 48824}

\begin{abstract}
The pulsar emission mechanism in the gamma-ray energy band is poorly understood.
Currently, there are several models under discussion in the pulsar community.
These models can be constrained by studying the collective properties of a sample of pulsars, which became possible with the large sample of gamma-ray pulsars discovered by the Fermi Large Area Telescope (Fermi-LAT).
In this paper we develop a new experimental multi-wavelength technique to determine the beaming factor $\left( f_\Omega \right)$ dependance on spin-down luminosity of a set of GeV pulsars.
This technique requires three input parameters: pulsar spin-down luminosity, pulsar phase-averaged GeV flux and TeV or X-ray flux from the associated Pulsar Wind Nebula (PWN).
The analysis presented in this paper uses the PWN TeV flux measurements to study the correlation between $f_\Omega$ and $\dot{E}$.
The measured correlation has some features that favor the Outer Gap model over the Polar Cap, Slot Gap and One Pole Caustic models for pulsar emission in the energy range of 0.1 to 100 GeV, but one must keep in mind that these simulated models failed to explain many of the most important pulsar population characteristics.
A tight correlation between the pulsar GeV emission and PWN TeV emission was also observed, which suggests the possibility of a linear relationship between the two emission mechanisms.
In this paper we also discuss a possible mechanism to explain this correlation.
\end{abstract}


\section{Introduction}
Currently, there are several GeV pulsar models being discussed in the pulsar community.
These models can be constrained either by studying individual pulsars in detail or alternatively using the collective properties of a sample of pulsars.
The large sample of pulsars discovered by Fermi Large Area Telescope (Fermi-LAT) provides a good place to study the collective properties of GeV pulsars.\\

The GeV luminosity as a function of other pulsar parameters is a fundamental quantity which models must predict.
However, the potential utility of pulsar luminosity is limited by two factors: inability to measure the beaming factor $\left( f_\Omega \right)$ (also called the beam correlation factor) \citep{FOmega} and imprecise distance measurements.
The factor $f_\Omega$ provides the correction to extrapolate the observed phase-averaged flux from the earth line-of-sight to the full sky flux for a given beam shape.
It is an essential factor needed to convert observed fluxes to luminosity:
\begin{equation}
 L_P = 4\pi d^2 f_\Omega F,
\end{equation}
where $L_P$ is the luminosity, $d$ is the distance to the pulsar from earth and $F$ is the phase averaged flux measured at earth.
Since $f_\Omega$ is a model dependent parameter, luminosity calculations are also model dependent.
Therefore, one option to constrain GeV pulsar emission models is to use the collective properties of the luminosity distribution but the uncertainty on distance measurements degrades the accuracy of the luminosity distribution.  This issue can be resolved by studying the ratio of the flux from pulsars to that of their associated Pulsar Wind Nebulae (PWNe), which is a distance independent parameter.\\

This paper uses the ratio between a pulsar's GeV flux and the TeV flux from its associated PWN as the first application of this method.
Using this ratio we obtain the dependence of $f_\Omega$ on spin-down luminosity $\dot{E}$ for a sample of pulsars.
This allows us to compare the experimentally measured  dependence of $f_\Omega$ on $\dot{E}$ with the theoretical
expectation of four $\gamma$-ray pulsar gap models \citep{GammarayConstrain}. \\

\section{The sample of pulsars and their associated PWNe}\label{FermiGeVPulsars}
Recently the Fermi Large Area Telescope (Fermi-LAT) produced its second pulsar catalog \citep{SecondFermiLATPSRPaper} with 117 high-confidence $\gamma$-ray ( $\geq$ 0.1 GeV) pulsars.
In addition, TeV $\gamma$-ray observatories, such as Milagro, VERITAS and H.E.S.S., have measured TeV fluxes coming from PWNe.
From a literature survey we found fourteen GeV pulsars in the Fermi-LAT pulsar catalog for which the associated PWNe were also measured by TeV observatories.\\

Table \ref{tab:PSRandPWN} and \ref{tab:PSRandPWN2} summarizes the properties of these 14 objects.
This analysis uses the values of spin-down luminosity ($\dot{\textrm{E}}$), distance to the pulsar ($d$) and phase averaged flux in the energy range of 0.1-100 GeV ($G_{100}$) reported in the Fermi-LAT Second Pulsar Catalog.\\

\subsection{Discussion of TeV PWNe Measurements}\label{TeVMeasurements}
The integrated energy flux around 35 TeV of the associated PWNe $\left(F_{TeV}\right)$ are listed in column 3 of Table \ref{tab:PSRandPWN2}.
All Milagro TeV measurements in this column are derived from Table 1 in \cite{MilagroBSL}.
Hereafter, we will refer to \cite{MilagroBSL} as the Milagro 0FGL search.
In that publication the Milagro collaboration performed a targeted search for galactic sources in the Fermi Bright Source List, which is also known as 0FGL \citep{FermiBSL}.
The Milagro 0FGL search found TeV emission coincided with 14 Fermi bright sources.
Among these 14 sources, 9 have spatial associations with pulsars in the Fermi-LAT Second Pulsar Catalog.
The Milagro 0FGL search paper reported the differential photon flux at 35 TeV $\left( \frac{\textrm{dN}}{\textrm{dE}} \mid_{35~TeV} \right)$ assuming a Spectral Energy Distribution (SED) of E$^{-2.6}$ without a cutoff.
The authors argue that the flux calculated at 35 TeV has the least dependence of the calculated flux on the true spectrum.
Because Milagro does not report the full SED for these pulsars, but only the differential photon flux at 35 TeV, we calculated the integrated energy flux over a 1 TeV band around 35 TeV with a SED of E$^{-2.6}$ and used the Milagro flux uncertainties.\\

The Milagro 0FGL search paper mentioned that TeV emission might come from the pulsar and/or from the associated PWN.
However, it is very unlikely to get a significant contribution from the pulsar to the TeV flux measured by Milagro.
The best example of this is the Crab pulsar, which is the brightest TeV object measured by Milagro.
The VERITAS Collaboration \citep{VeritasCrabPSRPaper} observed pulsed $\gamma$-rays in the energy range of $\sim100$ GeV to $\sim200$ GeV.
The measured energy spectrum is well described by a simple power law, without a cut-off, by:
\begin{equation}
\frac{\textrm{dN}}{\textrm{dE}} = \left( 4.2 \pm 0.6_{stat} {\pm^{2.4}_{1.4}}_{syst} \right) \times 10^{-11} \left( \frac{\textrm{E}}{150~\textrm{GeV}} \right)^{-3.8 \pm 0.5_{stat} \pm 0.2_{syst}} ~~\textrm{TeV}^{-1}\textrm{cm}^{-2}\textrm{s}^{-1}
\end{equation}
An extrapolation of this energy spectrum gives a differential photon flux of $4.2 \times 10^{-20}~~\textrm{photons}~\textrm{TeV}^{-1}\textrm{cm}^{-2}\textrm{s}^{-1}$ at 35 TeV.
This is 0.003\% of the TeV flux observed coincident with the Crab pulsar by Milagro.
In addition, a theoretical model proposed in \cite{AharonianGeVPulsarmodel} predicts a sharp cut-off below $\sim$500 GeV, so
the extrapolated flux at 35 TeV from the pulsar might be even lower.
These considerations lead us to conclude that the TeV emissions observed coincident with pulsars come predominantly from their associated PWNe.\\

We performed a literature search for PWNe measured by Air Cherenkov telescopes in the TeV band.
We found H.E.S.S. SEDs for five other PWNe, which are also listed in Table \ref{tab:PSRandPWN} and \ref{tab:PSRandPWN2}: Crab, Vela, K3 in Kookabura, MSH 15-52, and G 21.5-0.9.
In order to be consistent with the Milagro measurements, we integrated the energy flux within a 1 TeV energy band around 35 TeV using the H.E.S.S. SEDs.
The SEDs of the Crab, Vela and MSH 15-52 PWNe were measured in the energy ranges of 440 GeV - 40 TeV, 550 GeV - 65 TeV and 250 GeV - 40 TeV, respectively.
Therefore, their integrated energy flux around 35 TeV can be obtained without any extrapolation.
However, the SEDs of K3 in Kookabura and G 21.5-0.9 PWNe were measured in the energy ranges of 200 GeV - 25 TeV and 150 GeV - 5 TeV, respectively.
		Therefore, their integrated energy fluxes obtained around 35 TeV by extrapolation of their SEDs:q1 might be an overestimate, if there is a cutoff below 35 TeV.\\

VERITAS has published SEDs of the Boomerang and CTA 1 PWNe.
In both cases the SEDs of these sources were measured in the energy range of 1-15 TeV.
Therefore,  as for K3 and G 21.5-0.9, the integrated energy flux obtained around 35 TeV by extrapolating the SEDs might be an over estimate, if there is a cutoff before 35 TeV.
For all H.E.S.S. and VERITAS measured PWNe, errors on the integrated flux are estimated by a standard Gaussian Monte Carlo propagation of the uncertainties of the SED fit, with the $16^{\textrm{th}}$ percentile as the lower error bar and the $84^{\textrm{th}}$ percentile as the upper.\\

There are two independent measurements for both Boomerang and the Crab.  In each case, the measurements agree within experimental errors and differ by less than a factor of two.
Both measurements for these PWN are shown in the following plots but we use their weighted average
when doing fits, which does not alter any of the conclusions in this analysis.

\begin{deluxetable}{ccccc}
%\tabletypesize{\scriptsize}
%\footnotesize
\centering
  \tablecaption{ Properties of a sample of GeV pulsars cataloged in the Fermi-LAT Second Pulsar Catalog.
   \label{tab:PSRandPWN}  }
\tablewidth{0pt}
  %\begin{tabular}{@{}cccccc}
  %\hline
  %\hline
\tablehead{
\colhead{Pulsar Name}&\colhead{Association}	&  \colhead{$\log_{10}( \dot{E}  $} & \colhead{$G_{100}$ ($10^{-11}$}	& \colhead{Distance}\\
  PSR		&  				                   & ( $10^{34}$ erg s$^{-1}))$				&  erg cm$^{-2}$ s$^{-1}) $ 	& kpc	\\ 
}

\startdata

J0007+7303	&	CTA 1			&	35.65		&	40.1$\pm$0.4			&1.4$\pm$0.3		\\
J0534+2200	&	0FGL J0534.6+2201$^\dagger$	&	38.64		&	129.3$\pm$0.8			&2.0$\pm$0.5	\\
--\textquotedbl--& 	Crab			& 	--\textquotedbl--&	--\textquotedbl--		& --\textquotedbl--	\\
J0631+1036	&	0FGL J0631.8+1034$^\dagger$	&	35.24		&	4.7$\pm$0.3			&1.0$\pm$0.2	\\
J0633+1746	&	0FGL J0634.0+1745$^\dagger$	&	34.52		&	423.3$\pm$1.2			&$0.2^{+0.2}_{-0.1}$\\
J0835-4510	&	Vela			&	36.84		&	906$\pm$2			&$0.29\pm0.02$		\\
\\											
J1420-6048	&	K3 in Kookabura		&	37.01		&	17.0$\pm$1.4			&5.6$\pm$0.9		\\
J1509-5850	&	MSH 15-52		&	35.71		&	12.7$\pm$0.7			&2.6$\pm$0.5		\\
J1833-1034	&	G21.5-0.9		&	37.53		&	5.9$\pm$0.5			&4.7$\pm$0.4		\\
J1907+0602	&	0FGL 1907.6+0602$^\dagger$ 	&	36.45		&	25.4$\pm$0.6			&3.2$\pm$0.3	\\
J1958+2846	&	0FGL 1958.1+2848$^\dagger$ 	&	35.53		&	9.1$\pm$0.4			&$<18.5$	\\
\\													
J2021+3651	&	0FGL 2020.8+3649$^\dagger$ 	&	36.53		&	49.4$\pm$0.8			&$10.0^{+2.0}_{-4.0}$\\
J2021+4026	&	0FGL 2021.5+4026$^\dagger$ 	&	35.06		&	95.5$\pm$0.9			&1.5$\pm$0.4		\\
J2032+4127	&	0FGL 2032.2+4122$^\dagger$ 	&	35.44		&	10.6$\pm$0.6			&3.7$\pm$0.6		\\
J2229+6114	&	0FGL 2229.0+6114$^\dagger$ 	&	37.35		&	25.3$\pm$0.4			&0.8$\pm^{+0.15}_{-0.20}$\\
--\textquotedbl--&	Boomarang		&--\textquotedbl--	&	--\textquotedbl--		&	--\textquotedbl--\\
\enddata

\tablecomments{
	    $G_{100}$ is the phase averaged flux of the pulsar GeV emission in the 0.1-100 GeV energy band.
}

\end{deluxetable} 

\begin{deluxetable}{ccc}
%\tabletypesize{\scriptsize}
%\footnotesize
\centering
  \tablecaption{ TeV flux of the associated PWNe of a sample of GeV pulsars cataloged in the Fermi-LAT Second Pulsar Catalog.
   \label{tab:PSRandPWN2}  }
\tablewidth{0pt}
  %\begin{tabular}{@{}cccccc}
  %\hline
  %\hline
\tablehead{
\colhead{Pulsar Name}&\colhead{Association}	&\colhead{$F_N$ $(10^{-14}$}\\
  PSR			&  				&  TeV s$^{-1}$ cm$^{-2})$\\ 
}

\startdata

J0007+7303	&	CTA 1				& 	${1.4^{+4.9}_{-1.1}}^a$ \\
J0534+2200	&	0FGL J0534.6+2201$^\dagger$	&	$5.4\pm0.3^b$\\
--\textquotedbl--& 	Crab				&	${2.3^{+6.0}_{-2.0}}^c$\\
J0631+1036	&	0FGL J0631.8+1034$^\dagger$	&	1.5 $\pm$ 0.4$^b$\\
J0633+1746	&	0FGL J0634.0+1745$^\dagger$	&	1.2 $\pm$ 0.4$^b$\\
J0835-4510	&	Vela				&	${16.4^{+9}_{-7}}^d$\\
\\											
J1420-6048	&	K3 in Kookabura			& 	5.4${\pm_{1.3}^{1.7}}^e$\\
J1509-5850	&	MSH 15-52			& 	6.2${\pm_{0.8}^{0.9}}^f$\\
J1833-1034	&	G21.5-0.9			&	0.99${\pm^{1}_{0.6}}^g$\\
J1907+0602	&	0FGL 1907.6+0602$^\dagger$ 	&	3.9 $\pm$ 0.5$^b$\\
J1958+2846	&	0FGL 1958.1+2848$^\dagger$ 	&	1.1 $\pm$ 0.3$^b$\\
\\													
J2021+3651	&	0FGL 2020.8+3649$^\dagger$ 	&	3.6 $\pm$ 0.3$^b$\\
J2021+4026	&	0FGL 2021.5+4026$^\dagger$ 	&	1.2 $\pm$ 0.3$^b$\\
J2032+4127	&	0FGL 2032.2+4122$^\dagger$ 	&	2.1 $\pm$ 0.3$^b$\\
J2229+6114	&	0FGL 2229.0+6114$^\dagger$ 	&       2.3 $\pm$ 0.4$^b$\\
--\textquotedbl--&	Boomarang			&      ${1.3^{+15}_{-1.2}}^h$ \\
\enddata

\tablecomments{
	    $F_N$ is the integrated energy flux of the PWN TeV emission in the 34.5-35.5 TeV energy band.
}
\tablenotetext{a}{[VERITAS Measurement. Energy flux derived by extrapolating the SED.] Reference \cite{CTA1Veritas}}
\tablenotetext{b}{[Milagro Measurement] Reference \cite{MilagroBSL}}
\tablenotetext{c}{[H.E.S.S. Measurement] Reference \cite{CrabHESS}}
\tablenotetext{d}{[H.E.S.S. Measurement] Reference \cite{VelaXHESS}} 
\tablenotetext{e}{[H.E.S.S. Measurement. Energy flux derived by extrapolating the SED.] Reference \cite{KookaburaHESS}}		
\tablenotetext{f}{[H.E.S.S. Measurement] Reference \cite{MSHHESS}}		
\tablenotetext{g}{[H.E.S.S. Measurement.  Energy flux derived by extrapolating the SED.] Reference \cite{GTwentyOneHESS}}		
\tablenotetext{h}{[VERITAS Measurement.  Energy flux derived by extrapolating the SED.] Reference \cite{CTA1Veritas}} 
\end{deluxetable}

\section{Method}
\label{Sec:Method}
The ratio of the pulsar GeV luminosity ($L_P$) to the luminosity of the associated PWN ($L_N$) can be written in terms of the corresponding pulsar and PWN Flux:

\begin{equation}
 \begin{split}
   \frac{L_P}{L_N} &= \frac{4 \pi d^2 \fom G_{100}}{4 \pi d^2 F_N}\\
		     &= f_\Omega \times \frac{G_{100}}{F_N}
 \end{split}
\end{equation}

Taking their ratio cancels the distance but retains the beaming factor $\fom$, which can be written as:
\begin{equation}
\fom = \frac{L_p}{L_n} r
\label{Eq:LumRatio}
\end{equation}
where
\begin{equation}
r = \frac{F_N}{G_{100}},
\end{equation}
which is the observed flux ratio.
This relation is a mathematical identity which is valid for each individual pulsar and its associated PWN. This identity however can not be used to derive $\fom$ for a given pulsar, because $L_{P}$ is not measurable without $\fom$.
However, we can extract information on the $\edot$ dependence of $\fom$ for a selected group of pulsars by using models that predict the $\edot$ dependence of $L_P$ and $L_N$.

Many high energy pulsar models (e.g. \cite{HardingFirstPulsar, PolarCapHeating, OGModelByTakata, SlotGap}  and \cite{SlotGapRevised}) predict a power law relationship between $L_P$ and $\edot$.
\begin{equation}
L_P = k_P \cdot \edot^q
\label{Eq:PulsarEdotRelation}
\end{equation}
For a given pulsar, $k_P$ is independent of $\edot$, but depends on other pulsar properties such as the angle between the direction of the magnetic dipole axis and the rotation axis.  Both $k_P$ and the power $q$ are model-dependent.  Later in the paper, we will discuss the implications of different choices of $q$.

For PWNe \cite{MattanaXT} discussed the correlations between $\edot$ and PWN luminosity in TeV and X-ray energy bands.
Using H.E.S.S. measurements they showed that PWN TeV luminosity is not correlated with $\edot$.
This observation is consistent with the theoretical expectation, TeV photons are generated by the accumulated high-energy electrons in PWNe \citep{MattanaXT}.
Therefore, for a given ensemble of GeV pulsars we can choose a characteristic PWNe TeV luminosity $k_N$, independent of $\edot$.
%Substituting $L_N$ in equation \ref{Eq:LumRatio} by $k_N$ allows us to eliminate the distance uncertainties.
%The systematic errors due to using a characteristic value for $L_N$ is discussed in section \ref{Sec:Analysis}. }
\cite{MattanaXT} also showed that the X-ray luminosity vs $\edot$ distribution can be fitted into a power law model.
Therefore we can generalize the  X-ray luminosity vs $\edot$ distribution and TeV luminosity vs $\edot$ distribution as;
\begin{equation}
L_N = k_N \cdot \edot^m,
\label{Eq:PWNModel}
\end{equation}
where $m=0$ for TeV luminosity vs $\edot$ distribution.

Both of these energy bands are good candidates for applying our method.
The work presented in this paper uses PWN TeV luminosity.
Another analysis with PWN X-ray luminosity is in progress.

When we combine the model expectations in Equation \ref{Eq:PulsarEdotRelation} for pulsars and Equation \ref{Eq:PWNModel} for PWNe with Equation \ref{Eq:LumRatio}, we obtain $\fom$ for a specific pulsar $i$:
\begin{equation}
{\fom}_i = \left(\frac{k_P}{k_N}\right)_i \cdot r_i \cdot \edot_i^{(q-m)}\ \ .
\label{Eq:fomi}
\end{equation}
In log-log space we can rewrite this equation as:
\begin{equation}
\log{{\fom}_i} = \log{\left(\frac{k_P}{k_N}\right)_i} + \log{r_i} + (q-m)\log{(\edot_i)}.
\label{Eq:loglog}
\end{equation}

In this equation  $r_i$ and $\edot_i$ are measurable quantities, but the coefficients $k_P$ and $k_N$ are unknowns and vary pulsar to pulsar.
In this paper we do not intend to measure the $\fom$ of individual pulsars.
Instead we intend to obtain the $\fom$ dependence on  $\edot$ for an ensemble of GeV pulsars using $\hat{k_P}$ and $\hat{k_N}$, where $\hat{k_P}$ and $\hat{k_N}$ are typical values of $kp_i$ and $kn_i$ appropriate for our ensemble of pulsars.
We make this explicit by defining $d_i$ as the difference between typical values and the pulsar-dependent  constants:
\begin{equation}
d_i = \log \left(\frac{k_P}{k_N}\right)_i - \log \left(\frac{\hat{k_P}}{\hat{k_N}}\right).
\end{equation}
We can rewrite Equation \ref{Eq:loglog} with the parameter $d_i$ as,
\begin{equation}
\log{{\fom}_i}  - d_i = \log{\left(\frac{\hat{k_P}}{\hat{k_N}}\right)} + \log{r_i} + (q-m)\log{(\edot_i)}.
\end{equation}
Although $d_i$ is not measurable for individual pulsars, we can use this expression to obtain the dependence of an estimate of ${\fom}_i$ ($\hat{\fom}_i$) on $\edot_i$, where
\begin{equation}
\begin{split}
\log \hat{\fom}_i &= \log {\fom}_i - d_i\\
				     &= \log \left(\frac{\hat{k_P}}{\hat{k_N}}\right) + \log r_i + (q-m) \log{\edot_i}
\end{split}
\label{Eq:xyi}
\end{equation}
Thus $d_i$ is a correction factor between using typical values and the unknown pulsar-dependent values.  
We will estimate the magnitude of any such effects in section \ref{Sec:Analysis}.

This summarizes our method of extracting the $\edot$ dependence of $\fom$.  We now proceed to discuss our choices for the constants $q$, $m$, $\hat{k_P}$ and $\hat{k_N}$ in more detail, and examine how well data supports these choices.

\section{TeV PWN Measurements}\label{sec:PWN}

%We now consider the suitability of the TeV data for this analysis.\\

\begin{figure}
 \centering
 \includegraphics[width=0.75\textwidth]{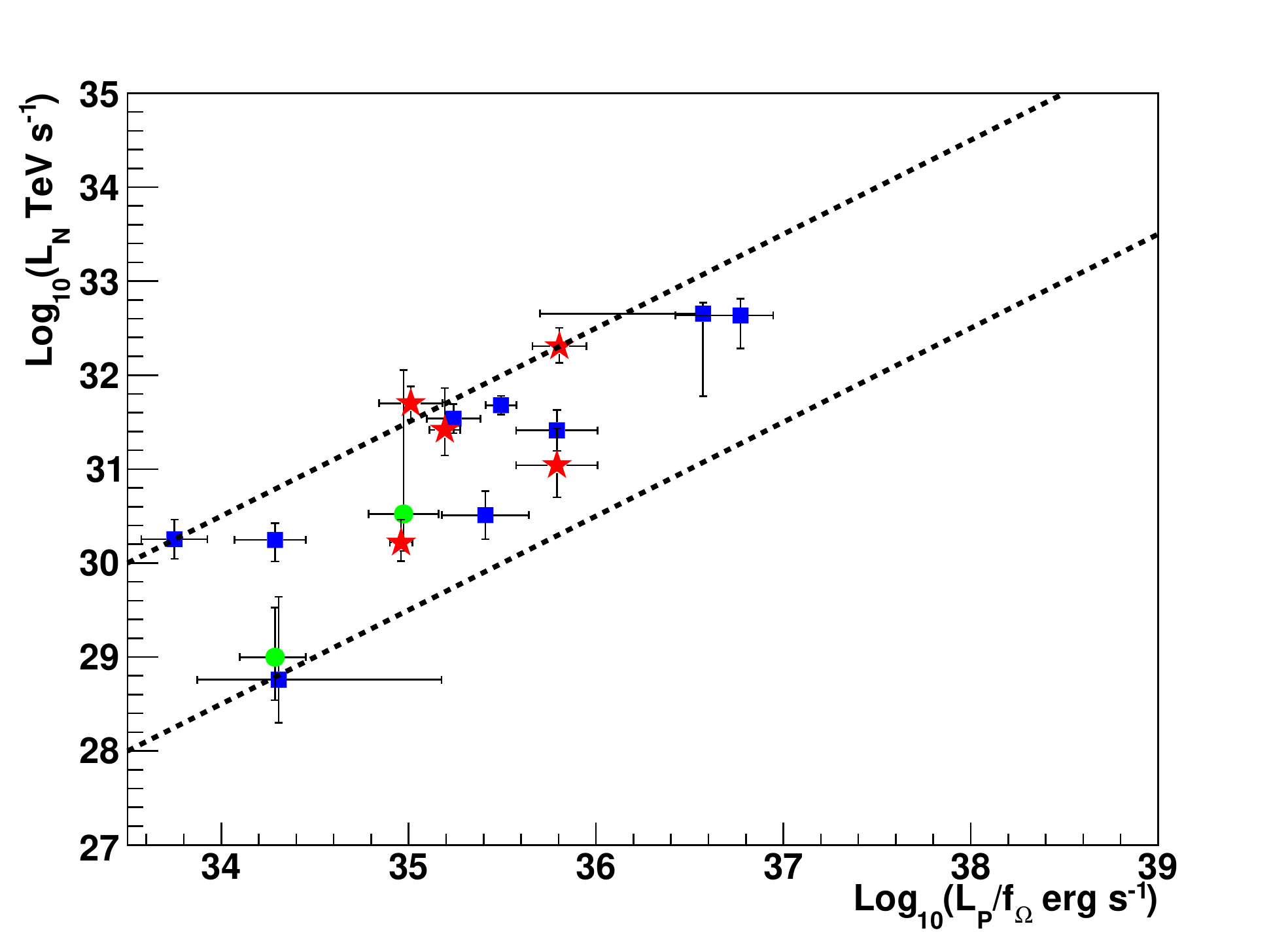}
  \caption{PWN TeV luminosity vs. Pulsar GeV luminosity normalized with respect to the beaming factor $\left( f_\Omega \right)$, $4 \pi d^2 G_{100}$.
	  Blue squares are measured by Milagro.
	  Red stars are measured by H.E.S.S.
	  Green circles are measured by VERITAS.
	  The two dotted lines have a slope of 1 but different arbitrary intercepts.
	  The linear correlation coefficient is R = 0.82.
	  The error bars are dominated by the distance measurement uncertainties.}
  \label{Fig:FluxCorrelation}
\end{figure}

First we consider whether pulsar and TeV PWN luminosity exhibit sufficient correlation to make it worthwhile to work with their ratio.
Figure~\ref{Fig:FluxCorrelation} shows the correlation between $L_P/\fom$ and $L_{N}$.  The blue squares are PWNe measured by Milgro, the red stars by H.E.S.S., and the green circles by VERITAS.
It appears that the PWNe measurements of the three experiments are generally consistent.
Four of the TeV measurements ($L_{N} =  34.3,35.0, 35.2, 35.8$) have been extrapolated from lower energy, but do not appear to be outliers to the general scatter of the distribution.
The large error bars on this plot are due to the distance uncertainties.
Also note that the TeV luminosity (y-axis) spans a larger range than the GeV (x-axis), making the error bars appear to be less important for TeV luminosities.
The dotted lines (drawn to guide the reader's eye) represent two lines of $L_{N} \propto \frac{L_{P}}{f_\Omega}$ with different arbitrary intercepts.
Even though the error bars are large a reasonable correlation is obtained with a linear correlation coefficient of R = 0.82, which
we judge to be sufficiently encouraging to proceed.

\begin{figure}
\centering
 \includegraphics[width=0.75\textwidth]{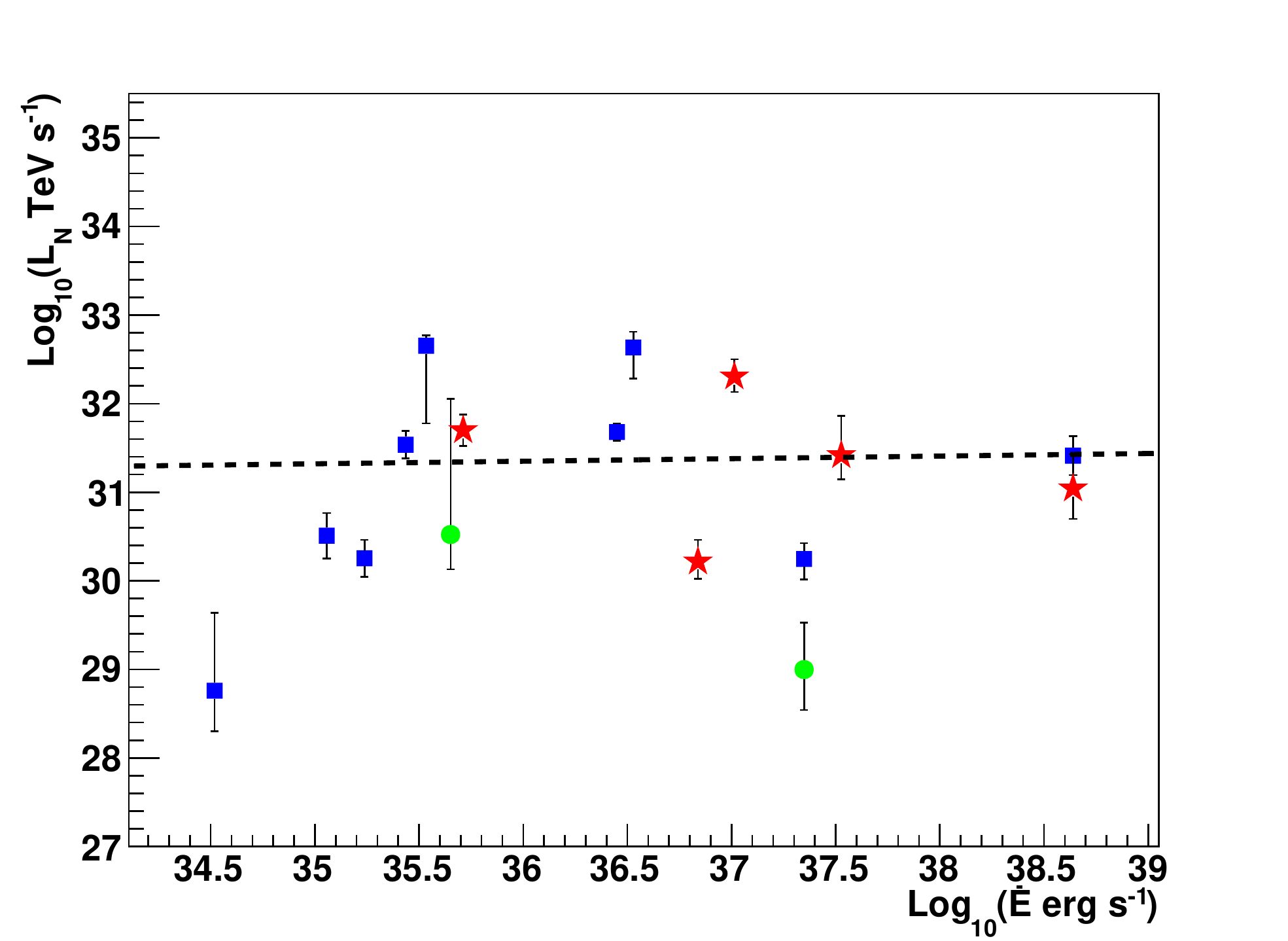}
\caption{ PWN TeV luminosity vs. spin-down luminosity $\dot{E}$ of the associated pulsar.
	  Blue squares are for PWNe measured by Milagro, red stars by H.E.S.S., and green circles by VERITAS.
    The distance uncertainty contributes significantly to the error bars.}
\label{Fig:TeVEdot}
\end{figure}

Next we examine the correlation between $L_{N}$ and $\dot{E}$.  
The $L_{N}$ vs $\dot{E}$ distribution for our PWN sample is shown in Figure~\ref{Fig:TeVEdot}.
Again we note that the uncertainty on the distance measurements contributes significantly to 
the luminosity error bars and that the extrapolated points ($\dot{E} = 35.7, 37.0, 37.3$ and 37.5 ) 
do not appear to be outliers.
This distribution has a linear correlation coefficient of 0.09.
The small linear correlation coefficient suggests that $L_N$ is not correlated with $\edot$ that concludes PWN TeV luminosity is not a function of $\edot$.
Therefore, one has to expect zero slope for the best fit linear fit for the data points.
The best fit linear fit for our data points has the slope of $0.03 \pm 0.06$, which is consistent with zero.
In summary, we argue that the observations are consistent with the theoretical expectation 
of no $\dot{E}$ dependence in the PWN TeV luminosity, which we discussed in section \ref{Sec:Method}.
Therefore, the model value m = 0 in Equation \ref{Eq:fomi}.
We fit a constant to the $\log L_N$ data, yielding $31.6 \pm .05$, 
and use this value for the model parameter $\log \hat{k_N} = 31.6$.

\section{GeV Pulsar Measurements}\label{sec:PSR}

Next we return to the power law model for the pulsed GeV emission $L_P$.
 We proceed with the analysis on the basis of $q = \frac{1}{2}$, because there are several high energy pulsar models which predict a power law index near $\frac{1}{2}$ (e.g. \cite{PolarCapHeating, OGModelByTakata, SlotGap}  and \cite{SlotGapRevised}), and further, because in this paper we compare our results with a sample of pulsars simulated using $L_P = k_P \edot^{\frac{1}{2}}$ as an underlying model assumption.
Later we will discuss the effect of $q$ on the best fit parameters in Equation \ref{Eq:model}.

Finally we can select a reasonable value of $\hat{k}_P$ by using Figure 9 of \cite{SecondFermiLATPSRPaper} which has an illustrative line for a $L_P = k_P \edot^{\frac{1}{2}}$ model with an additional constraint of $\fom=1$.
Taking a suitable point from the line, $\log \edot=39$ and $\log (L_p/\fom) = 36$, we find $\log \hat{k_P} = 16.5$.

\section{Analysis}
\label{Sec:Analysis}
\begin{figure}[H]
\centering
 \includegraphics[width=0.75\textwidth]{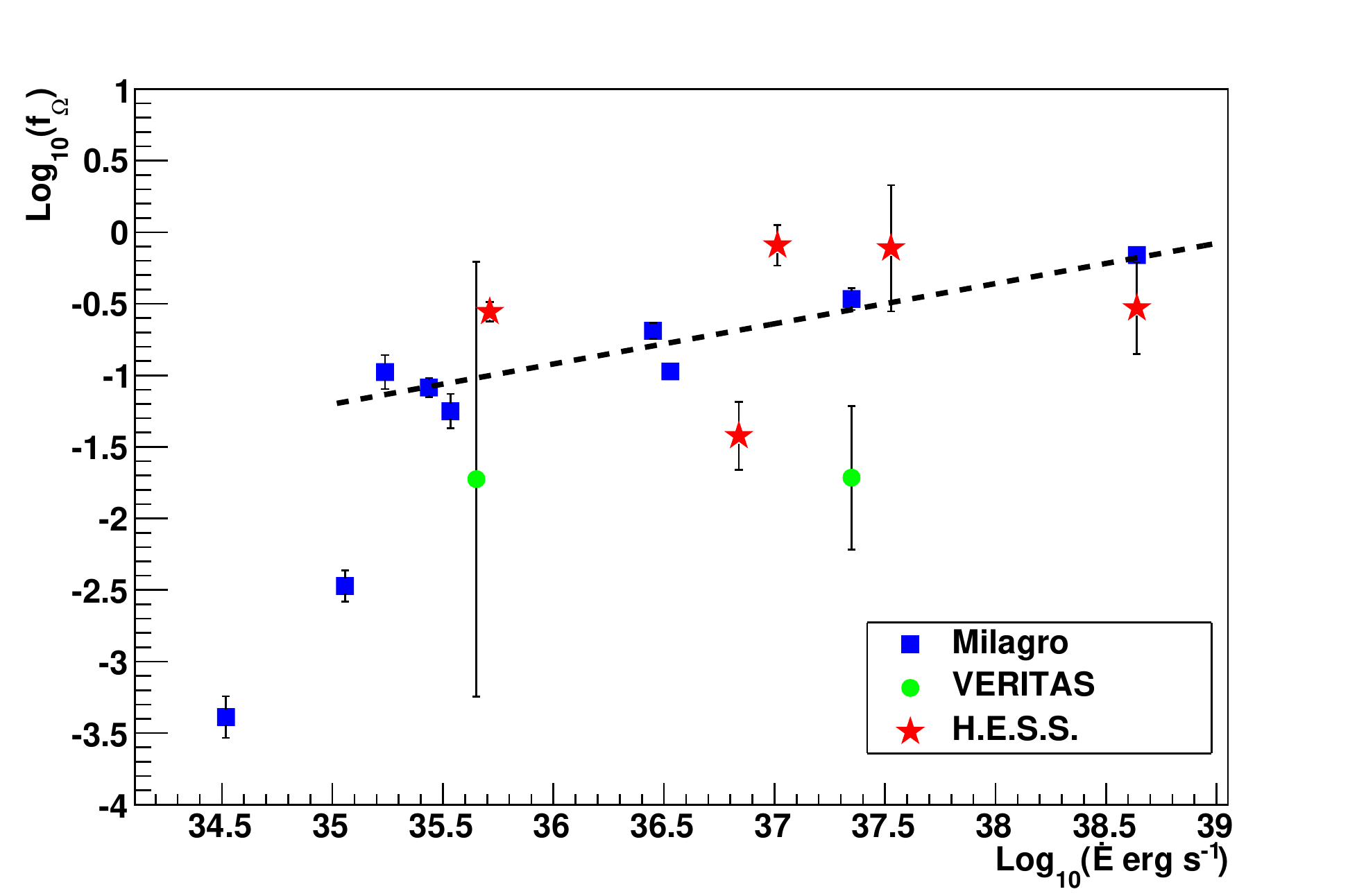}
  \caption{Experimentally obtained $f_\Omega$ vs. pulsar spin-down luminosity $\dot{E}$. Blue squares are PWNe measured by Milagro, red stars by H.E.S.S, and green circles by VERITAS. The best fit line for the pulsars with $\dot{E} > 10^{35.1}$ ergs s$^{-1}$ is shown as a dotted line.
  The slope of the best fit line is $0.28 \pm 0.03$.}
  \label{Fig:Ratio}
\end{figure}

As a summary of Sections \ref{sec:PWN} and \ref{sec:PSR}, we proceed with our analysis using model parameters $q - m = \frac{1}{2}$ and $\log_{10}(\hat{k}_p / \hat{k}_n) = -15.1$.
With these model parameters we can rewrite Equation \ref{Eq:xyi} as follows:
\begin{equation}
 \log_{10} \hat{\fom}_i = \log_{10}(r_i \cdot \dot{E}^\frac{1}{2}_i) - 15.1 
\end{equation}
The correlation between $ \log_{10}\hat{\fom}_i=y_i $ and $\log_{10} \edot_{i}=x_i$ is shown in Figure \ref{Fig:Ratio}.
It appears that above $\log \edot \approx 35 $, this distribution has a linear correlation with
\begin{equation}
\hat{y} = (-11.04 \pm 1.13) + (0.28 \pm 0.03) \cdot x_i,
 \label{Eq:Fit}
\end{equation}
where
\begin{equation}
\begin{split}
y_i          &= \log \hat{\fom}_i\\
             &= \log {\fom}_i - d_i \\
             &= \log \left(\frac{\hat{k_P}}{\hat{k_N}}\right) + \log r_i + (q-m) \log{\edot}_i\ \ \\
x_i &= \log {\edot}_i
\end{split}
\label{eq:modelfit}
\end{equation}
and $\chi^2$/NDF for this fit is 26.9/14.

Therefore, we can fit a phenomenological power law model for $\log \edot > 35 $,
\begin{equation}
\fom = a + b \cdot \log{\edot},
 \label{Eq:model}
\end{equation}
This power law model is motivated by \cite{GammarayConstrain}, and it also appears to be in agreement with the data. 
The $\chi^2$/NDF (= 26.9/14) for the fit in Equation \ref{Eq:Fit}  indicates deviations of ${\fom}_i$ from the fit line ($\hat{y_i}$) due to a combination of statistical measurement errors and systematical errors.
The systematic scatter can be estimated by adding a constant $\epsilon$ in quadrature with the statistical errors of each $y_i$, such that the chisquared per degree of freedom decreased to 1.0.
The resulting estimate is $\epsilon = 0.08$, which indicates $\sim 20\%$ systematic error.  
The slope of the fit with this added uncertainty is $0.27 \pm .06$, which is still inconsistent with zero.

The fit residuals represent the difference between the data points and the empirical fitting model:
\begin{equation}
y_i - \hat{y}_i = log {\fom}_i - d_i - \hat{y_i}
\end{equation}
Thus $\epsilon$ characterizes the scale of the differences among the estimated functional form (trend) $\hat{y}$, and the individual pulsar measurements, effectively considering the systematic deviations from the empirical power law for $\fom$ as well as the effects of removing $d_i$ from $\hat{\fom}$ (that is using the single pulsar-independent values of $k_P$ and $k_N$).  
We see that although these deviations represent the information about specific pulsars compared to the overall model (trend), the deviations of individual pulsars from the trend are not so large as to invalidate the model extraction of the trend, as this scale ($\epsilon = 0.08$) is notably smaller than the variation of $\hat{y}$ across the range of $\edot$.

\section{Discussion}\label{Sec::Discussion}

\subsection{Constraints on $\gamma$-ray pulsar gap models}

\begin{figure}
 \includegraphics[width=\textwidth]{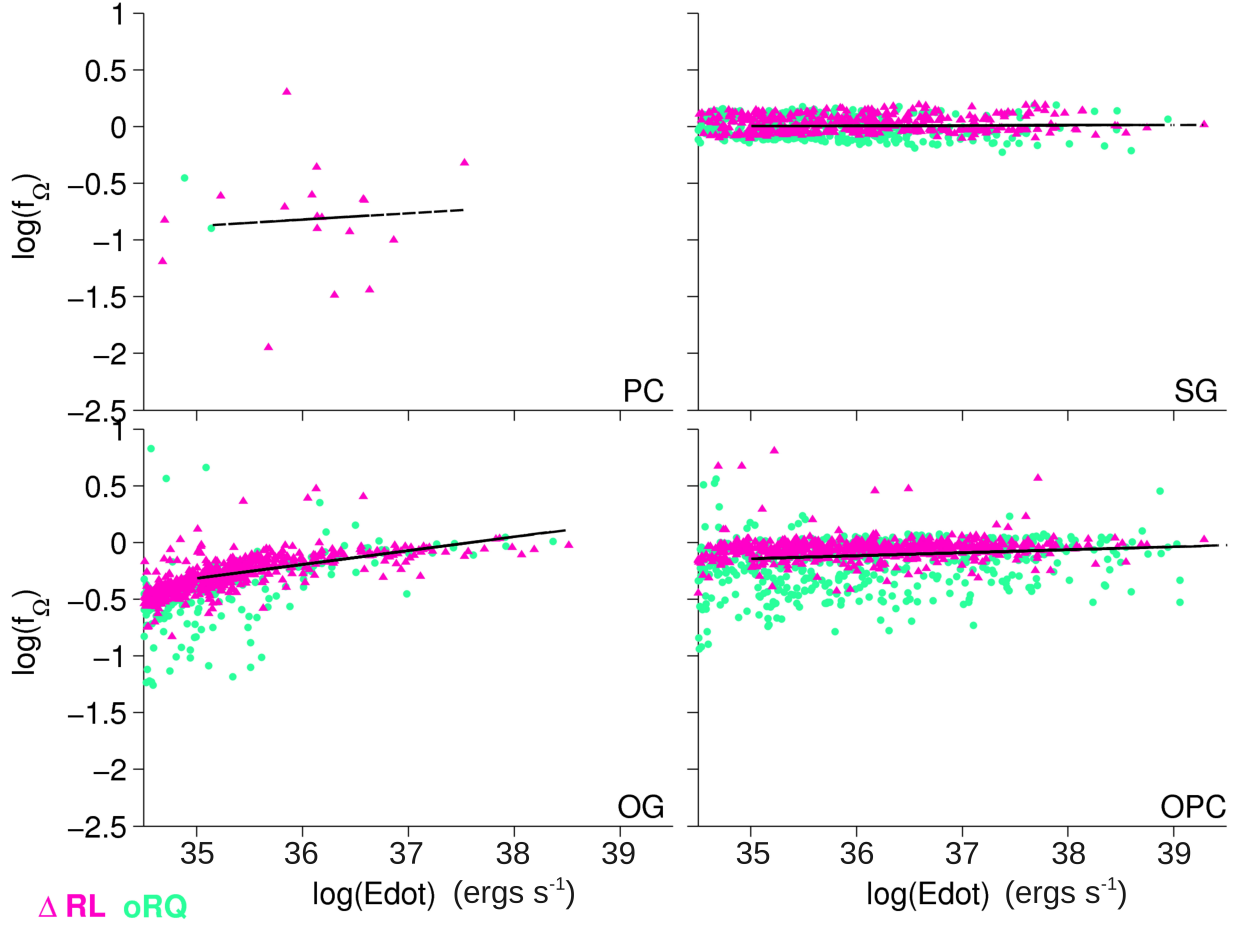}
\caption{The dependence of the beaming factor $f_\Omega$ on spin-down luminosity $\dot{E}$ for four models derived by \protect \cite{GammarayConstrain} a sample of simulated pulsars.
This plot was reproduced by M. Pierbattista with linear fits for pulsars above $\dot{E} = 10^{35}$ ergs s$^{-1}$.
For the original figure refer to \protect \cite{GammarayConstrain}.
Red and green markers refer to the radio-loud and radio-quiet pulsars, respectively.
Black lines refer to the best linear fits for the pulsars with $\dot{E} > 10^{35}$ ergs s$^{-1}$.
Slopes of the best fit lines are $m_{PC}  =  0.05 \pm  0.23$, $m_{SG} = 0.003 \pm  0.004$, $m_{OG} = 0.12 \pm  0.01$ and $m_{OPC} = 0.026 \pm 0.007$.}
\label{Fig:SimualatedFOmega}
\end{figure}

Recently, \cite{GammarayConstrain} studied four gamma-ray pulsar acceleration models.
They synthesized a pulsar population based on a radio emission model and four $\gamma$-ray pulsar gap models: Outer Gap (OG), Polar Cap (PC), Slot Gap (SG) and One Pole Caustic (OPC).
Their model simulations of the correlation between  $f_\Omega$ and $\dot{E}$ are shown in Figure~\ref{Fig:SimualatedFOmega}.
In all four model predictions, for pulsars with $\dot{E}>10^{35}$ ergs s$^{-1}$, $f_\Omega$ can be reasonably fit by a straight line in $\log - \log$ space and the best fits give the following slopes: $m_{PC}  =  0.05 \pm  0.23$, $m_{SG} = 0.003 \pm  0.004$, $m_{OG} = 0.12 \pm  0.01$ and $m_{OPC} = 0.026 \pm 0.007$.\footnote{These best fit parameters were provided to us by the authors of \cite{GammarayConstrain}.}
The slopes of the PC and SG models are consistent with zero, and the slope of the OPC is very small.
The OG model is the only model that predicts a positive slope that is not consistent with zero.
However, the small number of data points for the PC model have a large scatter and while the slope is consistent with zero,
the uncertainty on the slope is much larger than for the other models.\\

We can now compare our data points from Figure~\ref{Fig:Ratio} with the expectations from the four pulsar gap models.
First we note some gross characteristics in comparing the simulations to our data points.
Both the data and the models have $\log \fom$ near 0, and  the scatter of individual simulated pulsars about their trend line is not grossly different from the scatter in the data, despite the fact that the data extraction used single individual values of $k_N$ and $k_b$, while the simulation used  the full information about individual simulated pulsars.
This similarity is in agreement with the idea that the scatter due to $\epsilon_i$ is not so large as to lose all information about $\fom$.
These simulated data points show a tighter distribution of the simulated radio-loud pulsars to the best fit line above $\log_{10}(\edot) = 35$ while radio-quiet pulsars have a wider distribution.
Especially in the OG model radio-quiet pulsars deviate to low $f_\Omega$ values for $\log_{10}(\edot) < 35$.
We note that the radio-loud pulsars in our data also have a tighter distribution about the best fit line above $\log_{10}(\edot) = 35$ and the radio-quiet pulsars with  $\log_{10}(\edot) < 35$  deviate to low $f_\Omega$ values.
The authors of \cite{GammarayConstrain} also noted that the range of variation of their $\fom$ was less than that of their $L_P$ for the same range of $\edot$, at least for the SG and OG models.  We see a similar trend in our data.

The experimental $f_\Omega$ vs. $\dot{E}$ distribution has a non-zero slope of $b = 0.27 \pm 0.03$ for $\dot{E} > 10^{35}$ erg s$^{-1}$.
This would tend to disfavor the PC, SG and OPC models, despite the OPC providing the best overall agreement with Fermi-LAT pulsars among the models considered.
However, the slope in our data is over twice that expected by the OG model for  $\dot{E} > 35$ ergs s$^{-1}$.
Below that value of $\dot{E}$, the expected correlation between $f_\Omega$ and $\dot{E}$ for the OG model becomes more dispersed and the $f_\Omega$ distribution for a given $\dot{E}$ has a tail towards smaller $f_\Omega$ values, especially for radio quiet pulsars.
This feature is also consistent with the experimentally obtained $f_\Omega$ vs. $\dot{E}$ distribution.
The two radio quiet pulsars with $\edot$ below $10^{35}$ erg s$^{-1}$ (PSR J0633+1746 and PSR J2021+4026) have smaller $f_\Omega$ values compared to radio loud pulsars.
By considering both features above and below $\dot{E} \approx 10^{35}$ ergs s$^{-1}$, we can conclude that our data sample has some features which favor the OG model for  pulsar emission in the energy range 0.1-100 GeV over the SG and OPC models, though even the OG model does not quantitatively match our measurements.
However, we cannot reach any conclusions about the PC model.\\

The discrepancies with the models might be due to the systematic limitations or inadequacies of \cite{GammarayConstrain}'s simulations or biases in our data sample.
As the authors of \cite{GammarayConstrain} mention, results of all four simulated models are lacking pulsars visible by Fermi-LAT with $\dot{E} > 3 \times 10^{35}$ ergs s$^{-1}$ and characteristic age $< 100$ kyr, and over-predict the number of low $\dot{E}$ pulsars.
Furthermore they also mention that the simulated OG model in particular fails to explain many of the most important pulsar population characteristics, including the distributions of period, characteristic age, $L_P$ and $\edot$.
This could certainly affect their $f_\Omega$ vs. $\dot{E}$ distributions.
Therefore, we cannot provide tight constraints using this synthetic pulsar population.
However, a comparison of our results with an improved model simulation could provide tighter constraints.\\

A better sensitivity TeV survey such as HAWC promises to help our understanding of PWN visibility selection effects. 
Our estimates of $f_\Omega$ may also be biased if there is an unexpected residual dependence of the PWN TeV luminosity on $\dot{E}$ 
hidden in the TeV data scatter, or if there is hidden dependence on the selected TeV energy range.
The extracted slope of the $f_\Omega$ vs $\dot{E}$ distribution linearly depends on any hidden slope in the PWN TeV luminosity vs $\dot{E}$ distribution. 
The TeV energy range of a 1 TeV band around 35 TeV was chosen because of the Milagro data set.
Ideally, we would prefer to do this study with a more uniform sample of PWN TeV energy fluxes obtained around the 
inverse Compton peak of each individual PWN.
Using X-ray PWN luminosity instead of TeV luminosity would also offer complementary  visibility selection effects, 
to allow assessment whether such selection effects are important.\\

This result also depends on the uncertainties of the theoretically predicted correlation between GeV pulsar luminosity and spin-down luminosity, $L_{P} \propto \dot{E}^q$, $q=\frac{1}{2} + \delta$.
We made Figure~\ref{Fig:Ratio} for $L_{P} \propto \dot{E}^{\frac{1}{2}}$ ($\delta=0$), because it is natural in several pulsar models to have this relation \citep{FermiPSRPaper} and it is one of the underlying assumptions in \cite{GammarayConstrain}.
However, the slope of $\log_{10}(f_\Omega)$ vs. $\log_{10}(\dot{E})$ distribution depends linearly on $\delta$, as seen in Equation \ref{eq:modelfit}.
Therefore we can write that for $\dot{E} > 10^{35}$ erg s$^{-1}$ the slope of $\log_{10}(f_\Omega)$ vs. $\log_{10}(\dot{E})$ distribution is $0.28\pm0.03_{stat} + \delta$.
One can use this expression to determine the slope of $\log_{10}(f_\Omega)$ vs. $\log_{10}(\dot{E})$ distribution under different models.  For example, \cite{OGModelByTakata} discuss a high-energy emission from the outer gap that expects $L_{P} \propto \dot{E}^{\frac{5}{8}}$.
Under this model the slope $\delta = 0.125$ and slope becomes $0.405\pm0.03_{stat}$.
Another example is \cite{SlotGap}, which discuss a high-energy pulsar emission model that depends on the local magnetic field.
In the low magnetic field scenario \cite{SlotGap} expects $L_P \propto \dot{E}^{\frac{3}{7}}, \delta = -0.07$.
Under this model the slope becomes $0.21\pm0.03_{stat}$.

\subsection{Pulsar GeV emission to PWN TeV emission connection}

The correlation between the pulsar GeV emission and the PWN TeV emission shown in Figure~\ref{Fig:FluxCorrelation} leads one to suspect a common underlying cause for the two emission mechanisms.
One property relevant to both emissions is the electron-positron current of the pulsar wind $\left( I_{wind} \right)$.
The GeV energy flux from pulsars is thought to be directly related to the instantaneous value of $I_{wind}$, because the GeV pulsed emission from the magnetosphere is often thought to be produced by curvature emission by the most recently produced electron-positron population in the wind.
Since the luminosity is roughly proportional to the population of electrons and positrons, we can write:
\begin{equation}
 L_{PSR~GeV} \propto I_{wind}.
\end{equation}

Inside PWNe, TeV photons are often thought to be produced by the up-scattering of ambient photons by the relativistic electrons and positrons, known as inverse Compton radiation.
Therefore, $L_{N}$ should depend on the relativistic electron-positron population and the ambient photon population in the PWN.
However, for the relativistic electrons and positrons that produce TeV photons by inverse Compton scattering the typical cooling time is larger than the lifetime of pulsars \citep{MattanaXT}.
Therefore, the population of these electrons and positrons becomes proportional to the integral of $I_{wind}$ over the pulsar lifetime, instead of proportional to the instantaneous value of $I_{wind}$.
However, we could suggest a proportionality between the ambient photon field density $\left( \rho_{ph} \right)$ and $I_{wind}$.
There are two different ambient photon fields which could be relevant to the production of TeV $\gamma$-rays: photons from synchrotron radiation and far-infrared photons \citep{AtoyanCrabPaper}.
The density of synchrotron radiation photons in the x-ray energy band is roughly proportional to the density of the freshly injected pulsar wind \citep{MattanaXT}.
In addition far-infrared seed photons can be made by heating the pulsar wind, as described in Section 2.2 of \cite{AronsPESPaper}.
Therefore the $\rho_{ph}$ may be roughly proportional to $I_{wind}$.
If  $\rho_{ph}$ is proportional to $L_{N}$, that would yield:
\begin{equation}
 L_{N} \propto \rho_{ph} \propto I_{wind}.
\end{equation}
Hence,
\begin{equation}
 L_{N} \propto L_{PSR~GeV}.
\end{equation}
While these considerations are suggestive, a more detailed theoretical study is clearly needed to fully understand this correlation.

\section{Conclusion}\label{Conclusion}
We have developed a new multi-wavelength technique to study the collective properties of the GeV pulsar beaming factor $f_\Omega$ with respect to the pulsar spin down luminosity $\dot{E}$.
This technique uses a distance independent parameter, $\frac{F_N}{G_{100}}$, to obtain the correlation between  $f_\Omega$ and $\dot{E}$.
It allowed us to use the pulsars with poorly known distance measurements to study $f_\Omega$.
Using this technique we have experimentally obtained the $f_\Omega$ vs. $\dot{E}$ dependence for pulsar emission in the 0.1-100 GeV energy band.
Under the model assumptions of an $\dot{E}^{\frac{1}{2}}$ dependence of GeV pulsed emission but no $\dot{E}$ dependence of TeV PWN emission, we find a dependence of $f_\Omega$ on $\dot{E}$.
Our experimentally obtained correlation between $f_\Omega$ and $\dot{E}$ has some features which favor the theoretical $f_\Omega$ vs. $\dot{E}$ distribution of the OG model obtained by \cite{GammarayConstrain}.
However, this specific comparison is limited by the modeling uncertainties of \cite{GammarayConstrain}'s simulated pulsar sample.
Applying this same multi-wavelength method to X-ray data for PWNe is attractive, since it may have a more precisely measurable $\dot{E}$ dependence than the present TeV data.

Pulsar GeV emission and PWN TeV emission are correlated, with a linear correlation coefficient of R = 0.82, although TeV PWN emission has no correlation to $\dot{E}$.
This observed GeV to TeV correlation suggests the possibility of a linear relationship between $\gamma$-ray emission mechanisms in pulsars and TeV emission mechanisms in PWNe.
However, it is not possible to explain this linear relationship using the electron-positron populations of curvature radiation inside the magnetosphere and synchrotron radiation in the PWN.
An alternative possibility is a linear relationship between the ambient photon density $\left( \rho_{ph} \right)$ in the PWN and the pulsar wind current $\left( I_{wind} \right)$.
A more detailed theoretical study will be needed to fully understand this correlation.

In the near future, TeV experiments under development, such as HAWC, CTA, and Lhasso, will have greater sensitivity than Milagro.  The observed GeV to TeV luminosity correlation makes it likely that these observatories will detect PWNe associated with many more of the GeV pulsars Fermi has observed, leading to prospects of a higher-statistics and higher precision version of this analysis.

\acknowledgments
We would like to give our special thanks to the authors of \cite{GammarayConstrain}, particularly to Dr. M. Pierbattista for producing new fits for the simulated data and Prof. P. Gonthier for reading a draft of our paper and giving us constructive feedback.  We also thank Prof. K. Tollefson and Dr. J. Pretz for useful suggestions, and Dr. G. Sinnis and Prof. M. Voit for their encouragement to publish this work.  This work was supported by the National Science Foundation.


\begin{thebibliography}{}
\bibitem[Abdo et al.(2009a)]{MilagroBSL} Abdo, A.~A., Allen, B.~T., Aune, T., et al.\ 2009a, \apjl, 700, L127
\bibitem[Abdo et al.(2009b)]{FermiBSL} Abdo, A.~A., Ackermann, M., Ajello, M., et al.\ 2009b, VizieR Online Data Catalog, 218, 30046
\bibitem[Abdo et al.(2010)]{FermiPSRPaper} Abdo, A.~A., Ackermann, M., Ajello, M., et al.\ 2010, \apjs, 187, 460
\bibitem[Abdo et al.(2013)]{SecondFermiLATPSRPaper} Abdo, A.~A., Ajello, M., Allafort, A., et al.\ 2013, \apjs, 208, 17
\bibitem[Acciari et al.(2009)]{BoomerangVERITAS} Acciari, V.~A., Aliu, E., Arlen, T., et al.\ 2009, \apjl, 703, L6
\bibitem[Aharonian et al.(2005)]{MSHHESS} Aharonian, F., Akhperjanian, A.~G., Aye, K.-M., et al.\ 2005, \aap, 435, L17
\bibitem[Aharonian et al.(2006a)]{CrabHESS} Aharonian, F., Akhperjanian, A.~G., Bazer-Bachi, A.~R., et al.\ 2006a, \aap, 457, 899
\bibitem[Aharonian et al.(2006b)]{VelaXHESS} Aharonian, F., Akhperjanian, A.~G., Bazer-Bachi, A.~R., et al.\ 2006b, \aap, 448, L43
\bibitem[Aharonian et al.(2006c)]{KookaburaHESS} Aharonian, F., Akhperjanian, A.~G., Bazer-Bachi, A.~R., et al.\ 2006c, \aap, 456, 245
\bibitem[Aharonian et al.(2012)]{AharonianGeVPulsarmodel} Aharonian, F.~A., Bogovalov, S.~V., \& Khangulyan, D.\ 2012, \nat, 482, 507
\bibitem[Aliu et al.(2013)]{CTA1Veritas} Aliu, E., Archambault, S., Arlen, T., et al.\ 2013, \apj, 764, 38
\bibitem[Arons(1996)]{AronsPESPaper} Arons, J.\ 1996, \aaps, 120, 49
\bibitem[Atoyan \& Aharonian(1996)]{AtoyanCrabPaper} Atoyan, A.~M., \& Aharonian, F.~A.\ 1996, \mnras, 278, 525
\bibitem[Harding(1981)]{HardingFirstPulsar} Harding, A.~K.\ 1981, \apj, 245, 267
\bibitem[Harding \& Muslimov(2002)]{PolarCapHeating} Harding, A.~K., \& Muslimov, A.~G.\ 2002, \apj, 568, 862
\bibitem[Harding \& Muslimov(2003)]{SlotGapRevised} Harding, A.~K., \& Muslimov, A.~G.\ 2003, Pulsars, AXPs and SGRs Observed with BeppoSAX and Other Observatories, 121
\bibitem[H.~E.~S.~S.~Collaboration et al.(2007)]{GTwentyOneHESS} H.~E.~S.~S.~Collaboration, :, Djannati-Atai, A., et al.\ 2007, arXiv:0710.2247
\bibitem[Mattana et al.(2009)]{MattanaXT} Mattana, F., Falanga, M., G{\"o}tz, D., et al.\ 2009, \apj, 694, 12
\bibitem[Muslimov \& Harding(2003)]{SlotGap} Muslimov, A.~G., \& Harding, A.~K.\ 2003, \apj, 588, 430
\bibitem[Nolan et al.(2012)]{Fermi2FGL} Nolan, P.~L., Abdo, A.~A., Ackermann, M., et al.\ 2012, \apjs, 199, 31
\bibitem[Pierbattista et al.(2012)]{GammarayConstrain} Pierbattista, M., Grenier, I.~A., Harding, A.~K., \& Gonthier, P.~L.\ 2012, \aap, 545, A42
\bibitem[Romani \& Watters(2010)]{RomaniSimulation} Romani, R.~W., \& Watters, K.~P.\ 2010, \apj, 714, 810
\bibitem[Takata et al.(2010)]{OGModelByTakata} Takata, J., Wang, Y., \& Cheng, K.~S.\ 2010, \apj, 715, 1318
\bibitem[VERITAS Collaboration et al.(2011)]{VeritasCrabPSRPaper} VERITAS Collaboration, Aliu, E., Arlen, T., et al.\ 2011, Science, 334, 69
\bibitem[Watters et al.(2009)]{FOmega} Watters, K.~P., Romani, R.~W., Weltevrede, P., \& Johnston, S.\ 2009, \apj, 695, 1289






\end{thebibliography}
\end{document}